\documentclass[
 aip,
 amsmath,amssymb,
 reprint,%
]{revtex4-1}

\usepackage{graphicx}
\usepackage{dcolumn}
\usepackage{bm}

\usepackage[utf8]{inputenc}
\usepackage[T1]{fontenc}
\usepackage{mathptmx}
\usepackage{etoolbox}
\usepackage{siunitx}
\usepackage{nicefrac}
\makeatletter
\def\@email#1#2{%
 \endgroup
 \patchcmd{\titleblock@produce}
  {\frontmatter@RRAPformat}
  {\frontmatter@RRAPformat{\produce@RRAP{*#1\href{mailto:#2}{#2}}}\frontmatter@RRAPformat}
  {}{}
}%
\makeatother

\begin{document}

\preprint{AIP/123-QED}

\title[Cryogenic geometric anti-spring vibration isolation system]{Cryogenic geometric anti-spring vibration isolation system}
\author{L.Feenstra}
 \affiliation{Fakultät für Physik, Ludwig-Maximilians-Universität, Schellingstr. 4, Munich, 80799, Germany.}%
 \affiliation{Leiden Institute of Physics, Leiden University, P.O. Box 9504, 2300 RA Leiden,
The Netherlands}

 \email{L.Feenstra@lmu.de}
\author{S. Domínguez-Calderón}%
 \affiliation{Fakultät für Physik, Ludwig-Maximilians-Universität, Schellingstr. 4, Munich, 80799, Germany.}%
\author{K. van Oosten}
\affiliation{Leiden Institute of Physics, Leiden University, P.O. Box 9504, 2300 RA Leiden,
The Netherlands}
\author{H.S.M. Bohemen}
\affiliation{Leiden Institute of Physics, Leiden University, P.O. Box 9504, 2300 RA Leiden,
The Netherlands}
\author{T. Benschop}
\affiliation{Leiden Institute of Physics, Leiden University, P.O. Box 9504, 2300 RA Leiden,
The Netherlands}
\author{M. Brinkman}
\affiliation{Leiden Institute of Physics, Leiden University, P.O. Box 9504, 2300 RA Leiden,
The Netherlands}
\author{M. Li}
\affiliation{Leiden Institute of Physics, Leiden University, P.O. Box 9504, 2300 RA Leiden,
The Netherlands}
\author{E. Hennes}
 \affiliation{National Institute for Subatomic Physics, Nikhef Amsterdam, Science Park 105, 1098 XG Amsterdam, The Netherlands}
\author{R. Cornelissen}
 \affiliation{National Institute for Subatomic Physics, Nikhef Amsterdam, Science Park 105, 1098 XG Amsterdam, The Netherlands}
\author{B.J. Hensen}
 \affiliation{Leiden Institute of Physics, Leiden University, P.O. Box 9504, 2300 RA Leiden,
The Netherlands}
\author{A. Bertolini}
 \affiliation{National Institute for Subatomic Physics, Nikhef Amsterdam, Science Park 105, 1098 XG Amsterdam, The Netherlands}
\author{M.P. Allan}
 \affiliation{Fakultät für Physik, Ludwig-Maximilians-Universität, Schellingstr. 4, Munich, 80799, Germany.}%
 \affiliation{Leiden Institute of Physics, Leiden University, P.O. Box 9504, 2300 RA Leiden,
The Netherlands}

\date{\today}

\begin{abstract}
The combination of low temperature and low vibration levels is key for ultrasensitive sensing applications such as scanning probe microscopy, large-mass quantum mechanics, and gravitational wave detection. Unfortunately, closed-cycle cryostats using pulse tube or GM coolers introduce strong low-frequency vibrations starting at 1~Hz. Mass-spring systems allow passive isolation, but for low-frequency applications the required spring constants and masses become impractical. Blade-based geometric anti-spring systems are compact isolators that operate from sub-Hz frequencies, but have not been demonstrated at cryogenic temperatures. Here, we characterize a geometric anti-spring system tuned to operate at cryogenic temperatures. Our cryogenic filter uses radially arranged titanium blade springs whose effective spring constant can be tuned in-situ using a magnetic actuator. Our system achieves a vertical resonance frequency of 185 mHz at 7K, which allows reduction of vibrations at the problematic 1~Hz cooler frequency by an order of magnitude.
\end{abstract}

\maketitle

\section{Introduction}
\begin{figure}
    \centering
    \includegraphics{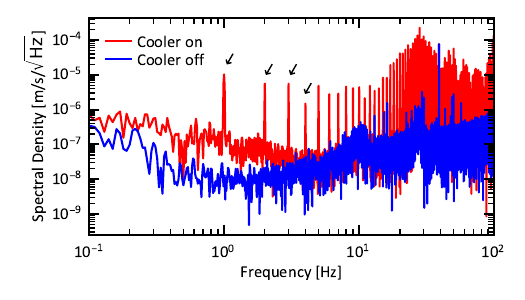}
    \caption{A comparison of vertical vibrations of the 4K-plate of our GM based  cryostat (see Fig 2 for a schematic), measured by a geophone. If the GM cooler is on, vibrations are stronger by orders of magnitude. The strongest noise is visible at a frequency of \SI{1}{Hz} and  harmonics thereof, the first four of which are indicated with arrows.}
    \label{fig:motivation}
\end{figure}

Scanning probe microscopy (SPM) \cite{battisti_definition_2018} and ultra-sensitive force sensing for quantum gravity experiments \cite{bose_spin_2017} both require ultra-low vibrations at cryogenic temperatures -- either to reduce noise in the tunneling current to a few fA \cite{battisti_definition_2018}, or to detect effects of gravity with sensitivities in the sub-fN regime \cite{fuchs_measuring_2024}. 
The same requirements apply to operational and future interferometer-based gravitational wave detectors, where the mirrors are suspended inside a cryogenic environment \cite{akutsu_vibration_2021, koroveshi_cryogenic_2023, utina_etpathfinder_2022}.
The challenge of achieving low vibrations is compounded by the growing popularity of closed-system cryostats, driven by helium shortage and correspondingly high prices \cite{noauthor_mineral_2025}.
These systems use pulse tube or GM coolers, which introduce very strong vibrations from pressure changes, typically around \SI{1}{Hz}, as can be seen in Fig.~\ref{fig:motivation}. 
Therefore, vibration attenuation especially at low frequencies is the key to realize ultra-sensitive experiments in closed-cycle cryostats.

Vibration attenuation can be achieved by both active and passive systems, which are often combined to optimize the attenuation properties.
In this paper, we focus entirely on the latter type. 
Passive vibration insulation conventionally consists of classical mass-spring systems, where the experiment is located on the mass.
Such mass-spring systems act as a low-pass filter for vibrations going from ground to mass, since vibrations at frequencies above the resonance frequency $\omega_0$ are damped as $\nicefrac{1}{\omega^2}$. 
This property implies that the performance of a mass-spring system can be optimized by minimizing the resonance frequency, given by $\omega_0 = \sqrt{{k}/{M}}$,
where $k$ is the spring constant and $M$ is the suspended mass.
Hence, by reducing the spring constant or increasing the mass, $\omega_0$ can be minimized, and the vibration attenuation at  frequencies $\omega ~> \omega_0$ is improved. 
Lowering the resonance frequency of conventional mass-spring systems has the problem of increasing system length, which can be in the order of meters for sub-Hz frequencies. In cryostats such length scales are not viable as space is generally limited.

To overcome the limitations of conventional mass-spring systems, we turn towards geometric anti-spring (GAS) filters to achieve resonance frequencies around a few \SI{100}{mHz} in a cryogenic environment.
These systems, first proposed by Bertolini et al. \cite{bertolini_seismic_1999}, are brought close to a mechanical instability by a compressive force perpendicular to the direction of the mass movement.
This so-called anti-spring effect results in a greatly reduced effective spring constant, as we describe in more detail later.

GAS filters have been realized at room temperature in gravitational wave detectors KAGRA and VIRGO \cite{akutsu_vibration_2021, koroveshi_cryogenic_2023, beker_seismic_2012, kumar_status_2016}, operating at resonance frequencies of a few 100~mHz.
As this is far below the operating frequency of closed-system cryostats, GAS filters are a promising candidate to attenuate pulse tube noise.
This possibility has been explored by Eßer et al. \cite{eser_ultra-high_2024} and the company Four Nine Design \cite{deshpande_demonstration_2025}, who both used a GAS filter at room temperature to suspend a cryogenic STM chamber via a thin wire. 
The experiment is then connected to the cryostat with thermal connections, which is a challenge as they  introduce  external vibrations to the experiment. 

In this paper, we present a GAS system that is designed to operate at cryogenic temperatures as a platform for ultra-low vibration cryogenic experiments. 
We choose to keep the whole system at base temperature, to keep the system simple and reduce vibrational short cuts due to thermalization wires.
The goal of this paper is to demonstrate that geometric anti-springs can be tuned an operated in a cryogenic environment, paving the way for GAS-based cryogenic vibration isolation.

We start the paper with an overview of the GAS and introduce the main equations governing the system. 
Starting from conventional mass-spring systems, a linear model is developed to demonstrate the mechanics and attenuating properties of GAS filters.
Then we describe the experimental design of the GAS and the cryostat. 
Next, we test the performance of our system and explain the measurement procedures. 
Finally, we identify the next steps to realizing cryogenic GAS filters for ultra-sensitive measurements.

\section{Theory of geometric anti-spring systems}
We start by developing the theoretical model for the dynamics of GAS filters, mainly describing them as tunable conventional mass-spring systems. 

\subsection{Conventional mass-spring systems}
Conventional mass-spring systems consist of a spring, fixed to some base at one end, and to a freely hanging mass on the other. 
The experiment will be on the mass, the source of the vibrations will be the base, and the goal is to reduce vertical vibrations of the mass above a certain frequency.

The system is characterized by the spring constant $k$ as well as the suspended mass $M$ and spring mass $m\ll M$.
The spring has a loaded length $l$.
We further include viscous damping, described by $\gamma$, and structural damping of the spring, described by the loss angle $\phi$.
We restrict motion to the vertical $y$-axis and define $y_0(t)$ and $y(t)$ as the respective vertical coordinates of the base and mass w.r.t. some inertial frame.

The equation of motion of such a mass-spring system is
\begin{equation}\label{Eq:eq of motion}
    M\ddot{y} = -k(1+i\phi)\left((y-(y_0-l)\right) - \alpha m\ddot y_0 - \gamma \dot{y}\text{.}
\end{equation}
The term $\alpha m \ddot y_0$ describes the center-of-percussion effect, arising in case of non-aligned forces on both spring ends.
In the case of a coil spring, this term vanishes as $\alpha = 0$.
Because it is non-zero in the case of a GAS \cite{stochino_improvement_2007}, we will include the term here.

We then introduce an outside source of vibrations to the spring base.
By Fourier transform of Eq.~\ref{Eq:eq of motion}, we obtain the transfer function $H_y(\omega)$, defined as the ratio between the mass' and the spring base's vertical motion, also called the filter's transmissibility.
The transfer function is given by:

\begin{equation}\label{Eq:transfer function}
    H_y := \frac{Y(\omega)}{Y_0(\omega)} = \frac{\omega_0^2(1+i\phi) + \alpha \frac{m}{M}\omega^2}{\omega_0^2(1+i\phi) - \omega^2+i\frac{\gamma}{M}\omega}\text{,}
\end{equation}
where $Y(\omega)$ and $Y_0(\omega)$ are the respective Fourier transformations of $y(t)$ and $y_0(t)$, and $\omega_0=\sqrt{\nicefrac{k}{M}}$. 

The low-pass filtering properties become clear when considering the limit of a massless undamped spring, $m=0, \gamma=0, \phi=0$.
At $\omega\ll\omega_0$, Eq.~\ref{Eq:transfer function} reduces to $H_y=1$, hence mass and base move in unison. 
$\omega\approx \omega_0$ results in an amplified motion of the mass as $H_y$ diverges, and $\omega\gg\omega_0$ leads to the desired damping of $\nicefrac{1}{\omega^2}$.
Thus, as seen in Eq.~\ref{Eq:transfer function} the vibration attenuation performance of the mass spring system is optimized by minimizing $\omega_0$.
For conventional mass-spring systems, reduction of $k$ or increasing of $M$ is cumbersome and competes with spatial constraints, effectively limiting resonance frequencies to around \SI{1}{Hz}. 
For example, a linear mass-spring system with a resonance frequency of 0.5~Hz shows a spring extension under load of $\nicefrac{Mg}{k} = \nicefrac{\omega_0^2}{g}\approx \SI{1}{m}$.
A GAS spring, tuned to such a frequency occupies only a small fraction of  vertical space, thanks to its non-linear spring behavior

\subsection{Geometric anti-spring systems}
Next, we will discuss the theory of GAS systems, by constructing a model of our GAS linear springs.
We show that their effective spring constant can be tuned to extremely low values, reaching resonance frequencies well below 1~Hz. 

Our GAS uses compressed blade springs placed radially symmetric around a key stone, from which a mass $M$ can be suspended (Fig. \ref{fig:system_overview}a).
Using the radial symmetry, we can characterize the system in two dimensions by considering a cross section of two opposing blades, as indicated in Fig. \ref{fig:system_overview}b.
In the following, we treat the horizontal and vertical force components separately and constrain movement to the vertical direction.

\begin{figure}
    \centering
    \includegraphics[width=0.95\linewidth]{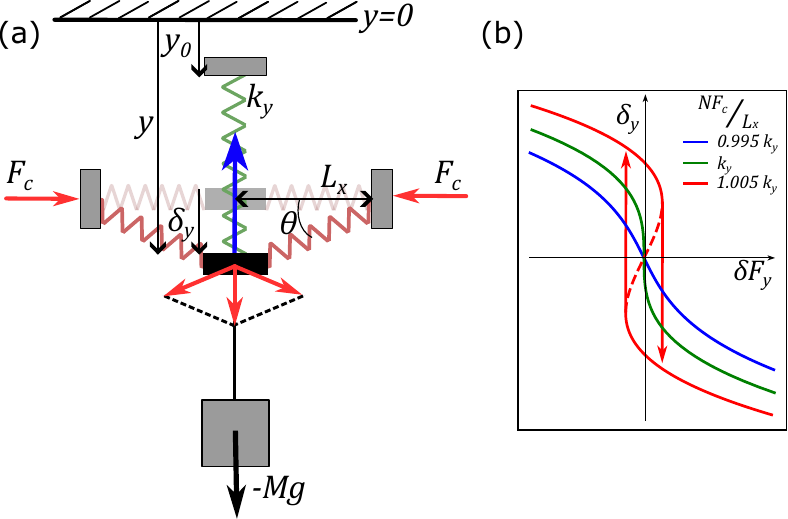}
    \caption{a) To develop the linear model for GAS, the vertical and compressive horizontal force components are separated. The vertical force is represented as a simple coil spring. The horizontal forces are transferred to the key stone via compressed springs. At the working height, the compressed springs are horizontal and the compressive forces cancel each other out. Adapted from Blom \cite{blom_seismic_2015}. b) S-curves following Eq. \ref{Eq:k_eff} for different compressive forces $F_c$. At $\delta y = 0$, all curves obtain minimal effective stiffness $k_\text{eff}$. The critical green curve has exactly $k_\text{eff}(\delta y) = 0$. The red curve represents the bistable state, where the minimal stiffness is negative.}
    \label{fig:linear_model}
\end{figure}

In the model of Fig. \ref{fig:linear_model}a, the vertical force component is obtained from a linear coil spring with stiffness $k_y$, carrying the payload.
The compressive horizontal component $F_c$ acts on the keystone through symmetrically configured linear springs. 
We define the working height as the situation in which the compressive forces act purely horizontally on the key stone, canceling each other out.
In this model, this corresponds to the compressed springs being horizontal.

Suppose the system is at rest with the key stone at the working height and is subject to a displacement $\delta_y$ from working height (See Fig.~\ref{fig:linear_model}a).
The vertical spring counteracts the deviation, while the compressed springs do the opposite, pushing the key stone further from the working point, hence the name anti-spring.
The net vertical force $\delta F_y=F_y-Mg$ in this configuration is

\begin{align}\label{Eq:LinModForce}
    \delta F_y &= -k_y \delta_y + NF_c \sin(\theta)  \\  
    &= -\left(k_y - \frac{NF_c}{\sqrt{L_x^2+\delta_y^2}}\right) \delta_y\text{,}
\end{align}

where $k_y$ is the spring constant for the vertical spring and $N$ refers to the number of blade springs in the GAS.
In this two-dimensional model, ${N=2}$.
The angle $\theta$ and distance $L_x$ can be inferred from Fig. \ref{fig:linear_model}a.

Based on Eq.~\ref{Eq:LinModForce} we approximate the effective spring constant for $\delta_y\ll L_x$ by differentiating $\delta F_y$ w.r.t. $\delta_y$
\begin{equation}\label{Eq:k_eff}
    k_\text{eff}(\delta_y)  \approx k_y - \frac{NF_c}{\sqrt{L_x^2+\delta_y^2}}\text{.}
\end{equation}

From this equation it becomes clear that the spring constant can be modified by tuning the compressive force $F_c$.

The goal of GAS filters is to tune them to their working point, where $F_c$ is chosen so that $k_\text{eff}$ is sufficiently low, but positive around the working height $\delta_y=0$. 
This configuration ensures optimal filtering performance.

If the springs are compressed beyond their working point, $k_\text{eff}<0$ for some range around $\delta_y=0$.
Within this range, there are two equilibrium positions symmetrically around $\delta_y=0$. 
Under sufficiently large forces, the mass can jump between the top and the bottom position. 

Experimentally, we extract the effective spring constant from measurements of the  so-called S-curve. 
This curve shows the equilibrium position as a function of load $M$. 
The effective spring constant is inversely proportional to the slope of the S-curve, having its minimum at $\delta_y = 0$.
Fig. \ref{fig:linear_model}b shows theoretical curves for three different values of $F_c$, where in the bistable regime, the curve is multi-valued.

The stiffness $k_\text{eff}$ can be used to describe the GAS transfer function using Eq. \ref{Eq:transfer function}.
For the GAS, the center-of-percussion effect leads to saturation of the $\nicefrac{1}{\omega^2}$-attenuation. 
$\alpha$ characterizes the strength of the effect, and is dependent on the blade geometry.
Stochino et al. \cite{stochino_improvement_2007} proposed a method to compensate the center-of-percussion effect, lowering the saturation level significantly.

\section{Design of a cyrogenic GAS and cryostat}

For our cryogenic vibration attenuation system, we chose a monolithic GAS design that is well-established in gravitational wave detectors \cite{akutsu_vibration_2021, beker_seismic_2012}.
It consists of quasi-triangular blade springs, arranged in a radial geometry (Fig \ref{fig:system_overview}a), clamped between a base and the key stone, from where the mass is suspended.

An exact 1 dimensional model of this geometry has been worked out by Cella et al. \cite{cella_monolithic_2005}, but the linearized model presented above describes the key characteristics.

\subsection{Blade design}
The most important considerations when designing such a GAS system concern the (i) shape, (ii) material, and (iii) size of the blade.
The combination of these three determines both the load that one can suspend from the filter and the lowest resonance frequency.
\begin{figure*}
    \centering
    \includegraphics[width = \linewidth]{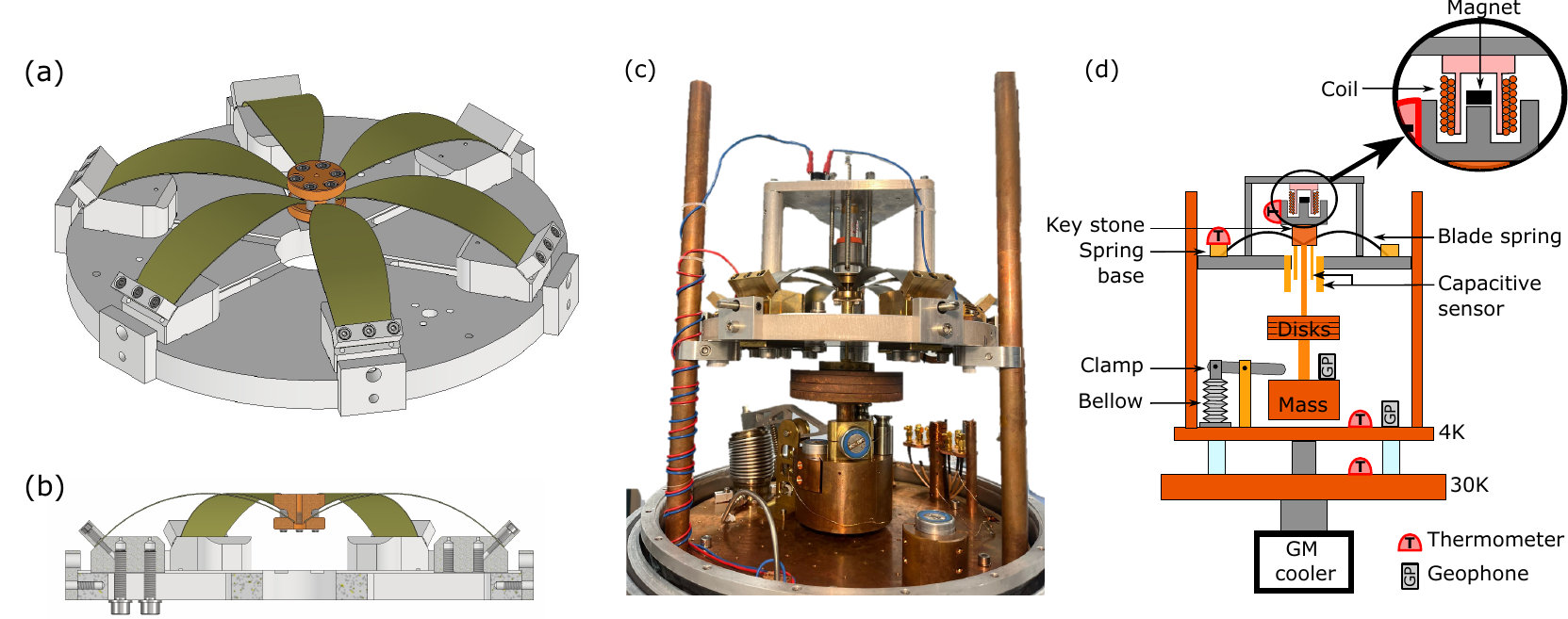}
    \caption{An overview of the presented cryogenic geometric anti-spring system is shown. \textbf{a)} A model of our monolithic GAS system, showing the six blade springs mounted on the spring plate. Notice the radially configured slits that alow tuning of the spring compression. \textbf{b)} Owing to the symmetry of the system, the system can be characterized by regarding a cross section. In this cross sectional view, the blade springs can be seen to be approximately uniformly bent, approaching the desired constant curvature under load. \textbf{c)} A photo of the experimental apparatus is shown. The bottom copper plate is the 4K-plate. The spring plate, hosting the GAS blades, is located directly above the 4K plate. \textbf{d)} A schematic cross-sectional view demonstrates the main components of the experimental apparatus.}
    \label{fig:system_overview}
\end{figure*}

(i) The shape of the blades determines the bending profile under load. 
As the resonance frequency $\omega_0\propto M^{-\nicefrac{1}{2}}$, it seems natural to design a blade shape that maximizes the load it can carry before exceeding the material's yield strength.
Furthermore, the saturation level of the $\nicefrac{1}{\omega^2}$ decay is decreased by reducing the ratio $\nicefrac{m}{M}$ (See Eq.~\ref{Eq:transfer function}).
Load maximization for limited blade size is achieved by the so-called constant-curvature solution \cite{blom_seismic_2015, cella_monolithic_2005}, where the blade bends uniformly when loaded.
However, as shown by Cella et al. \cite{cella_monolithic_2005}, this blade shape will not be able to reach the desired zero stiffness. 
They propose a shape that does reach zero stiffness, while approximating the constant curvature solution.

(ii) To maximize the load further, it is advisable to select a blade material with a high yield strength, such as titanium, (maraging) steel or beryllium copper - a choice that is also influenced by the cryogenic properties of the materials, as discussed below.

(iii) The size of the GAS blades also plays an important role in the filter performance. 
From the detailed model in \cite{cella_monolithic_2005}, one finds that the resonance frequency $\omega_0\propto L^{-\nicefrac{1}{2}}$, where $L$ is the length of the blade. 
In a cryostat, this poses a serious challenge, as space inside a cryostat is generally limited.

Our blades are designed based on these three considerations. 
The blade profile $W$ follows a fifth-order polynomial as a function of $\nicefrac{s}{L}$, where $0\leq s\leq L$ is the curvilinear coordinate along the blade.
$W$ is given by:
\begin{equation}
    W\left(\frac{s}{L}\right)=W_\text{ref} \sum_{i=0}^5 c_i\left(\frac{s}{L}\right)^i\text{,}
\end{equation}
where the coefficients $c_0 = 0.8317, c_1 = 0.9328, c_2=-1.9631, c_3=5.8534, c_4 = -12.22, c_5=6.6757$. $W_\text{ref}$ characterizes the width of the blade. In our system the dimensions are $W_\text{ref} = \SI{33.55}{mm}$ and $L=\SI{133.64}{mm}$. These were chosen to maximize the blade length and carrying strength, given the limited space inside the cryostat. 
The blades are made of \SI{0.6}{mm} thick grade 5 titanium (Ti-6Al-4V), which has a yield strength exceeding \SI{1}{GPa}.
In their working point, the springs are compressed, such that they depart from their base at an angle of $45^\circ$ and reach the key stone at an angle of $33^\circ$ with respect to the horizontal plane (Fig.~\ref{fig:system_overview}b).

\subsection{A compact, fast turn-over cryostat}
We load our  GAS-system into a home-built, fast turn-over cryostat, shown in Figs. \ref{fig:system_overview}c, e, based on a design of the Steele group at Delft University of Technology \protect\footnote{The drawings will be uploaded online soon.}. 
This home-built cryostat was specifically designed for this experiment. 
The requirements were (i) fast turn-over time and (ii) an easily accessible cold plate, with a base temperature below $\sim10$ K.

(i) The fast turn-over time was achieved by the use of a simple structure where the GM-cooler is directly connected to the base plate. We use a GM-cooler of type RDK-408D2 which has \SI{40}{W} of cooling power at \SI{40}{K}, and \SI{1}{W} at \SI{4}{K} to reach base temperature in under \SI{24}{hours}. 
Heaters on both stages and at the GAS base have a total power of \SI{450}{W}, allowing a warm-up within a few hours.

(ii) The  accessibility of the 4K-plate was achieved by placing the cooler and all feed-throughs on the bottom of the system, leaving the top position free for the 4K-plate (see Fig.~\ref{fig:system_overview}c, e).
The cylindrical vacuum and radiation shields connect to the rest of the cryostat at the height of the 4K-plate, making the entire cold space freely accessible from all directions.
The cold space has a diameter of 35 cm, and a height of 40 cm.

Three copper pillars are mounted on the 4K plate, to support the plate hosting the blade springs, such that the GAS height over the 4K plate can be chosen freely.
This height mainly depends on the suspended mass, which is located below the spring plate and is connected through a hole in the plate center.
This connection consists of a bronze rod, rigidly connected to the mass, which can be screwed into the key stone. 

If the mass is only thermalized via the titanium blade springs, the turn-over time increases drastically.
This issue can be resolved by clamping the suspended mass to the 4K-plate.
The clamp (see Figs. \ref{fig:system_overview}c, d) consists of a lever extending to the floating mass, connected to a vertical bellow at the other side. 
By regulating the He-gas pressure in the bellow between 0.5 and 1.5 bar, the bellow extends or retracts vertically.
In the extended state, the lever arm presses the mass to the 4K-plate, providing thermalization of this otherwise floating mass when cooling down or warming up.

\section{Performance}
The first step in testing our system is to tune the compression of the spring blades to achieve the minimal effective spring constant, or equivalently stiffness, at the working point.
After reaching this working point, we characterize the system and its performance.

\begin{figure}
    \centering
    \includegraphics{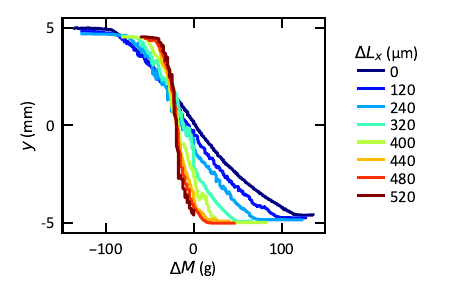}
    \caption{An iterative tuning process for our GAS filter based on recording S-curves, position-load diagrams. Initially, the compression is too low, so it is increased by shifting the spring bases inward by $\Delta L_x$ until reaching a near-vertical slope. The voice coil current is converted to a mass difference $\Delta M$.}
    \label{fig:tuning}
\end{figure}

\subsection{Tuning the system}
The optimal compression is found by iteratively changing the compression based on measured S-curves, as illustrated in Fig.~\ref{fig:tuning}.
The goal is to maximize the slope, or equivalently, minimize the stiffness at the working point.
If the measured stiffness is too high, one should increase the compression. 
In case of bistability, compression should be decreased.

The vertical position of the mass is recorded using a lock-in amplifier to read out a linear capacitive position sensor.
The sensor consists of two concentric, partially overlapping brass rods, that are electrically isolated from their environment and each other.
The wider one is connected to the spring plate, the narrower one to the key stone.
The rod connecting the key stone and the suspended mass runs through the center of the tubes. 
If the mass moves up and down, the overlap between the two tubes changes, leading to a change in capacitive value.

To tune the load on the filter, we can simply change the $\sim\SI{8}{kg}$ of suspended mass, consisting of a cylindrical copper block by adding copper disks and lab weights.  
For in-situ fine tuning, we equipped the system with a vertically oriented voice coil (see Fig.~\ref{fig:system_overview}d) which slightly pushes the mass up or down, corresponding with 85 mg/mA. 
To record S-curves in-situ, we sweep the coil current, thus sweeping the filter load.
From these curves, we extract the spring constant to tune the compression accordingly.

We can tune the system by moving the base of each spring individually, controlling the bending profile of the blades
Each base consists of two brass blocks between which the spring is clamped, as shown in Fig.~\ref{fig:system_overview}a, b.
These bases are clamped onto the spring plate with vertical screws mounted through radially configured slits in the spring plate.
On the outside of the spring plate, each spring base has a corresponding vertical tangential plate, with a fine-threaded screw (pitch 0.5 mm/turn) through it.
This screw pushes the base block inward and compensates the outward force from the blade spring, keeping the base in position.
By adjusting the fine screw, base position can be tuned with an accuracy of 20 $\mu$m, corresponding to 1/24th screw turn. 
In tuning procedure, as the base is moved inward, $F_c$ is increased and $L_x$ is decreased, thus decreasing $k_\text{eff}$ according to Eq. \ref{Eq:k_eff}. 

After tuning at room temperature, the cryogenic working point load is estimated based on the linear relation between the blades' force and its Young's modulus \cite{cella_monolithic_2005}.
For grade 5 titanium, they increase by $\sim10\%$ from ambient to cryogenic temperatures \cite{naimon_elastic_1974}, causing our mass to rest on the 4K-plate at ambient conditions.
Over the same temperature range, the optimal compression also changes due to different thermal expansion coefficients of the blades and spring plate. 
The thermal expansion difference \cite{marquardt_cryogenic_2001} between our titanium blades and aluminum spring plate of $2.5 \cdot10^{-3}$ corresponds to $\sim250\ \mu\text{m}$ spring base movement.
Therefore, we lower the compression by \SI{250}{\um} or $\nicefrac{1}{2}$ fine screw turn to approach optimal cryogenic compression.
With this estimated tuning as a starting point, we repeat the iterative process, this time recording S-curves at cryogenic temperatures and warming up to change the compression.
\begin{figure*}
    \centering
    \includegraphics[width = \linewidth]{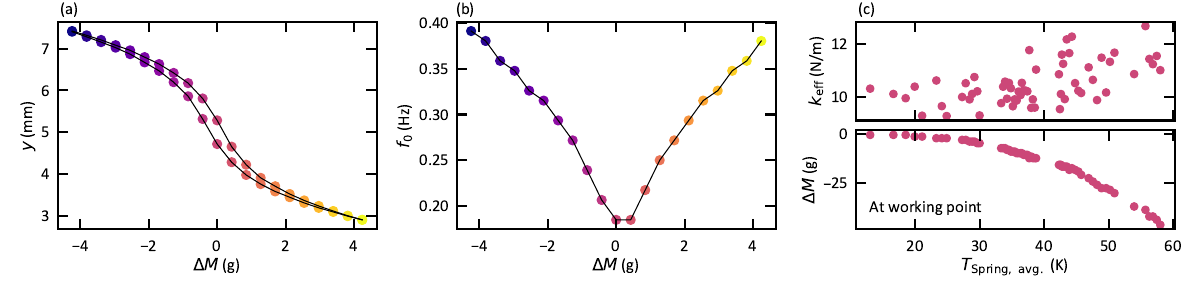}
    \caption{Characterization of the GAS filter. \textbf{a)} The S-curve of the tuned filter. Note that the zero-point $y=0$ is chosen differently from the data presented in Fig. \ref{fig:tuning}. b) For the tuned filter, the resonance frequency is recorded as a function of the applied load. We record a minimal frequency of \SI{185}{mHz}. c) As a function of temperature, the slope at the working point, shown in the top panel, undergoes slight changes due to the different thermal expansion coefficients of the blades and the base. Much more pronounced is the decreasing working point load due to the reducing Young's modulus, shown in the bottom panel. 
    }
    \label{fig:HO-data}
\end{figure*}

\subsection{Characterization of the working point}
With this iterative process we manage to reach near-zero stiffness at cryogenic temperatures, as we characterize in Fig.~\ref{fig:HO-data}.
In Fig.~\ref{fig:HO-data}a we show an S-curve at $T=\SI{7}{K}$, highlighting the need for the voice coil to achieve the sub-gram load tuning precision in-situ required for reaching the working point.
In this S-curve we also determined the resonance frequency as the load is changed, observing a minimal resonance frequency of 185~mHz.
This is shown in Fig.~\ref{fig:HO-data}b, demonstrating the strong dependence of the resonance frequency on the load: a few grams are enough to double the resonance frequency.

To investigate the temperature dependence of the working point, we use the heaters on the spring plate.
This introduces a temperature gradient to the springs, which can be of the order of tens of Kelvins, due to the poor heat conduction of titanium.
In Fig.~\ref{fig:HO-data}c, we plot the working point stiffness and load as a function of the average temperature of both spring ends.
Over the investigated temperature range, the stiffness changes only on the order of 10\%, the working point load however decreases by tens of grams. 
Hence, the performance of a GAS system is stable below $T\approx \SI{60}{K}$, as long as the coil can carry the current needed to reach the working point.



To measure the transmissibility of the filter, we use geophones, unidirectional accelerometers of the type Geospace GS-ONE-LF. 
These geophones act as a second-order high pass filter, with resonance frequency \SI{4.5}{Hz}, which we compensate with a home-built, analog second-order correction filter, as demonstrated in e.g. \cite{gilyazov_development_2023}.
This combination gives velocity response that can be considered linear for the scope of this work, providing a signal of $5.9\ \nicefrac{\text{V}}{\text{mm s}^{-1}}$ for $f>\SI{100}{mHz}$.

\begin{figure}
    \centering
    \includegraphics[width = \linewidth]{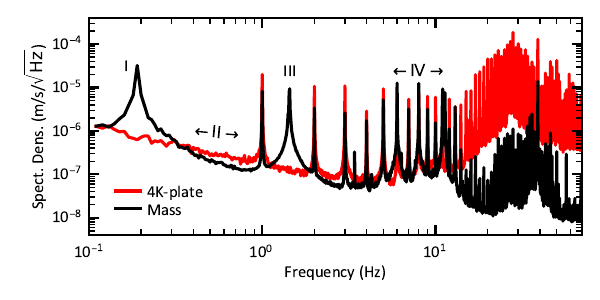}
    \caption{The absolute vibrations as recorded by the geophones at the cryogenic working point, averaged by Fourier transforming a 2000~s measurement in 100~s bins spaced at 50~s intervals. We indicate four different regimes. I) the vertical resonance at $f_0=\SI{185}{mHZ}$. II) The GAS provides attenuation of vertical vibrations. III) The pendulum mode at 1.45~Hz corresponds to our 10~cm pendulum length. IV) Around 10~Hz, apparent resonance peaks can be observed that may be attributed to modes breaking some symmetry of the system.}
    \label{fig:spectrum}
\end{figure}

We compare the vertical vibrations of the 4K-plate and the suspended mass to determine the attenuation properties of our GAS filter.
We obtain an average spectral density by Fourier transforming a 2000~s time series in bins of 100~s spaced at 50~s intervals.
This spectrum is shown in Fig.~\ref{fig:spectrum}.

In the mass spectrum, the vertical resonance is just below 200~mHz and the following damping can be identified below 1~Hz.
Around 1.45 Hz, there is a second mode, which we attribute to a pendulum mode.
This corresponds to a pendulum with a length of around 10~cm, which matches the distance between the center of mass and the key stone, which acts as pivot.
Around 10~Hz, further modes can be identified, whose origin is less clear. 
One could be a torsional mode in which the mass rotates back and forth in the horizontal plane, putting torsional load on the blades. 
Around 10~Hz we also observed rolling modes around the combined center of mass of the main mass and the tuning disks, putting a slight asymmetric load on the springs.

As Fig.~\ref{fig:spectrum} demonstrates, these additional modes at frequencies above the vertical resonance limit the damping performance of our GAS system.
It does however demonstrate that a GAS system can be tuned to a cryogenic working point with a vertical resonance well below 1~Hz, paving the way to realization of a GAS-based vibration isolation system for cryogenic experiments.

\section{Conclusions}
In closed-system cryostats, there is strong \SI{1}{Hz} noise present, which limits the feasibility of scanning probe microscopy or ultra-sensitive force measurements in such systems.
Starting from designs for gravitational wave observatories \cite{akutsu_vibration_2021, beker_seismic_2012}, we have shown the realization of a geometric anti-spring, that is designed for operation inside a cryostat.

In our system, we reduced the dimensions of the GAS filter and have included solutions for readout and mass fine tuning in situ at cryogenic temperatures.
This small GAS filter can be tuned to a vertical resonance frequency below \SI{200}{mHz}.
In the vibration spectrum, further modes can be identified as a pendulum and rotational modes, which should be damped before the low resonance frequency of the GAS filter can be used to optimally dampen vibrations for sensitive cryogenic experiments such as scanning probe microscopy.

Currently, we only use a single GAS, providing little attenuation for vibrations in the horizontal plane. This may be resolved by adding inverted pendulums, which can also reach extremely low resonance frequencies \cite{akutsu_vibration_2021, beker_seismic_2012}.
Another promising route is to investigate anti-springs in other geometries, to provide in-plane as well as vertical filtering \cite{allan_system_2025}.
In both approaches, active vibration isolation system to reduce the amplitude of the vertical resonance could be used to improve the system further. 
The compensation of the center-of-percussion effect as described in \cite{stochino_improvement_2007} could further improve the performance.

\section*{Acknowledgements}
This work was supported by the European Research Council Consilidator Grant (Project PairNoise).

\section*{Author declarations}
\subsection*{Conflicts of Interest}
The authors have no conflicts to disclose.

\subsection*{Author contributions}
\textbf{L. Feenstra}: Conceptualization (equal); Data curation (equal); Formal analysis (equal); Investigation (equal); Methodology (equal); Project administration (equal); Software (equal); Supervision (equal); Validation (equal); Visualization (equal); Writing – original draft (equal); Writing – review \& editing. \textbf{S. Dom\'inguez-Calder\'on}: Data curation (equal); Formal analysis (equal); Investigation (equal); Methodology (equal); Software (equal); Validation (equal); Writing – review \& editing. \textbf{K. van Oosten}: Investigation (equal); Methodology (equal); Writing – review \& editing. \textbf{H.S.M. Bohemen}: Investigation (equal); Methodology (equal); Writing – review \& editing. \textbf{T. Benschop}: Conceptualization (equal); Formal analysis (equal); Investigation (equal); Methodology (equal); Supervision (equal); Validation. \textbf{M. Brinkman}: Investigation (equal); Methodology (equal); Project administration (equal); Validation. \textbf{M. Li}: Data curation (equal); Formal analysis (equal); Investigation (equal); Methodology (equal); Project administration (equal); Supervision (equal); Validation. \textbf{E. Hennes}: Conceptualization (equal); Methodology (equal); Writing – review \& editing. \textbf{R. Cornelissen}: Conceptualization (equal); Methodology (equal); Writing – review \& editing. \textbf{B.J. Hensen}: Conceptualization (equal); Funding acquisition (equal); Methodology (equal); Project administration (equal); Supervision (equal); Validation (equal); Writing – review \& editing. \textbf{A. Bertolini}: Conceptualization (equal); Methodology (equal); Validation (equal); Writing – review \& editing. \textbf{M.P. Allan}: Conceptualization (equal); Funding acquisition (equal); Methodology (equal); Project administration (equal); Supervision (equal); Validation (equal); Writing – original draft (equal); Writing – review \& editing.

\section*{Data Availability Statement}
The data that support the findings of this study are available from the corresponding author upon reasonable request. 

\section*{References}
\bibliographystyle{aipnum4-1}
\bibliography{bibliography}

\begin{thebibliography}{20}%
\makeatletter
\providecommand \@ifxundefined [1]{%
 \@ifx{#1\undefined}
}%
\providecommand \@ifnum [1]{%
 \ifnum #1\expandafter \@firstoftwo
 \else \expandafter \@secondoftwo
 \fi
}%
\providecommand \@ifx [1]{%
 \ifx #1\expandafter \@firstoftwo
 \else \expandafter \@secondoftwo
 \fi
}%
\providecommand \natexlab [1]{#1}%
\providecommand \enquote  [1]{``#1''}%
\providecommand \bibnamefont  [1]{#1}%
\providecommand \bibfnamefont [1]{#1}%
\providecommand \citenamefont [1]{#1}%
\providecommand \href@noop [0]{\@secondoftwo}%
\providecommand \href [0]{\begingroup \@sanitize@url \@href}%
\providecommand \@href[1]{\@@startlink{#1}\@@href}%
\providecommand \@@href[1]{\endgroup#1\@@endlink}%
\providecommand \@sanitize@url [0]{\catcode `\\12\catcode `\$12\catcode `\&12\catcode `\#12\catcode `\^12\catcode `\_12\catcode `\%12\relax}%
\providecommand \@@startlink[1]{}%
\providecommand \@@endlink[0]{}%
\providecommand \url  [0]{\begingroup\@sanitize@url \@url }%
\providecommand \@url [1]{\endgroup\@href {#1}{\urlprefix }}%
\providecommand \urlprefix  [0]{URL }%
\providecommand \Eprint [0]{\href }%
\providecommand \doibase [0]{http://dx.doi.org/}%
\providecommand \selectlanguage [0]{\@gobble}%
\providecommand \bibinfo  [0]{\@secondoftwo}%
\providecommand \bibfield  [0]{\@secondoftwo}%
\providecommand \translation [1]{[#1]}%
\providecommand \BibitemOpen [0]{}%
\providecommand \bibitemStop [0]{}%
\providecommand \bibitemNoStop [0]{.\EOS\space}%
\providecommand \EOS [0]{\spacefactor3000\relax}%
\providecommand \BibitemShut  [1]{\csname bibitem#1\endcsname}%
\let\auto@bib@innerbib\@empty
\bibitem [{\citenamefont {Battisti}\ \emph {et~al.}(2018)\citenamefont {Battisti}, \citenamefont {Verdoes}, \citenamefont {van Oosten}, \citenamefont {Bastiaans},\ and\ \citenamefont {Allan}}]{battisti_definition_2018}%
  \BibitemOpen
  \bibfield  {author} {Battisti \emph {et~al.},\ }\href {\doibase 10.1063/1.5064442} {\bibfield  {journal} {\bibinfo  {journal} {Review of Scientific Instruments}\ }\textbf {\bibinfo {volume} {89}},\ \bibinfo {pages} {123705} (\bibinfo {year} {2018})}\BibitemShut {NoStop}%
\bibitem [{\citenamefont {Bose}\ \emph {et~al.}(2017)\citenamefont {Bose}, \citenamefont {Mazumdar}, \citenamefont {Morley}, \citenamefont {Ulbricht}, \citenamefont {Toroš}, \citenamefont {Paternostro}, \citenamefont {Geraci}, \citenamefont {Barker}, \citenamefont {Kim},\ and\ \citenamefont {Milburn}}]{bose_spin_2017}%
  \BibitemOpen
  \bibfield  {author} {Bose \emph {et~al.},\ }\href {\doibase 10.1103/PhysRevLett.119.240401} {\bibfield  {journal} {\bibinfo  {journal} {Physical Review Letters}\ }\textbf {\bibinfo {volume} {119}},\ \bibinfo {pages} {240401} (\bibinfo {year} {2017})}\BibitemShut {NoStop}%
\bibitem [{\citenamefont {Fuchs}\ \emph {et~al.}(2024)\citenamefont {Fuchs}, \citenamefont {Uitenbroek}, \citenamefont {Plugge}, \citenamefont {van Halteren}, \citenamefont {van Soest}, \citenamefont {Vinante}, \citenamefont {Ulbricht},\ and\ \citenamefont {Oosterkamp}}]{fuchs_measuring_2024}%
  \BibitemOpen
  \bibfield  {author} {Fuchs \emph {et~al.},\ }\href {\doibase 10.1126/sciadv.adk2949} {\bibfield  {journal} {\bibinfo  {journal} {Science Advances}\ }\textbf {\bibinfo {volume} {10}},\ \bibinfo {pages} {eadk2949} (\bibinfo {year} {2024})},\ \bibinfo {note} {publisher: American Association for the Advancement of Science}\BibitemShut {NoStop}%
\bibitem [{\citenamefont {Akutsu}\ \emph {et~al.}(2021)\citenamefont {Akutsu}, \citenamefont {Ando}, \citenamefont {Arai}, \citenamefont {Arai}, \citenamefont {Araki}, \citenamefont {Araya}, \citenamefont {Aritomi}, \citenamefont {Asada}, \citenamefont {Aso}, \citenamefont {Bae}, \citenamefont {Bae}, \citenamefont {Baiotti}, \citenamefont {Bajpai}, \citenamefont {Barton}, \citenamefont {Cannon}, \citenamefont {Cao}, \citenamefont {Capocasa}, \citenamefont {Chan}, \citenamefont {Chen}, \citenamefont {Chen}, \citenamefont {Chen}, \citenamefont {Chiang}, \citenamefont {Chu}, \citenamefont {Chu}, \citenamefont {Eguchi}, \citenamefont {Enomoto}, \citenamefont {Flaminio}, \citenamefont {Fujii}, \citenamefont {Fujikawa}, \citenamefont {Fukunaga}, \citenamefont {Fukushima}, \citenamefont {Gao}, \citenamefont {Ge}, \citenamefont {Ha}, \citenamefont {Hagiwara}, \citenamefont {Haino}, \citenamefont {Han}, \citenamefont {Hasegawa}, \citenamefont {Hatoya}, \citenamefont {Hattori}, \citenamefont {Hayakawa}, \citenamefont
  {Hayama}, \citenamefont {Himemoto}, \citenamefont {Hiranuma}, \citenamefont {Hirata}, \citenamefont {Hirose}, \citenamefont {Hong}, \citenamefont {Hsieh}, \citenamefont {Huang}, \citenamefont {Huang}, \citenamefont {Huang}, \citenamefont {Huang}, \citenamefont {Huang}, \citenamefont {Hui}, \citenamefont {Ide}, \citenamefont {Ikenoue}, \citenamefont {Imam}, \citenamefont {Inayoshi}, \citenamefont {Inoue}, \citenamefont {Ioka}, \citenamefont {Ito}, \citenamefont {Itoh}, \citenamefont {Izumi}, \citenamefont {Jeon}, \citenamefont {Jin}, \citenamefont {Jung}, \citenamefont {Jung}, \citenamefont {Kaihotsu}, \citenamefont {Kajita}, \citenamefont {Kakizaki}, \citenamefont {Kamiizumi}, \citenamefont {Kanda}, \citenamefont {Kang}, \citenamefont {Kawaguchi}, \citenamefont {Kawai}, \citenamefont {Kawasaki}, \citenamefont {Kim}, \citenamefont {Kim}, \citenamefont {Kim}, \citenamefont {Kim}, \citenamefont {Kim}, \citenamefont {Kimura}, \citenamefont {Kita}, \citenamefont {Kitazawa}, \citenamefont {Kojima}, \citenamefont
  {Kokeyama}, \citenamefont {Komori}, \citenamefont {Kong}, \citenamefont {Kotake}, \citenamefont {Kozakai}, \citenamefont {Kozu}, \citenamefont {Kumar}, \citenamefont {Kume}, \citenamefont {Kuo}, \citenamefont {Kuo}, \citenamefont {Kuromiya}, \citenamefont {Kuroyanagi}, \citenamefont {Kusayanagi}, \citenamefont {Kwak}, \citenamefont {Lee}, \citenamefont {Lee}, \citenamefont {Lee}, \citenamefont {Leonardi}, \citenamefont {Li}, \citenamefont {Lin}, \citenamefont {Lin}, \citenamefont {Lin}, \citenamefont {Lin}, \citenamefont {Lin}, \citenamefont {Liu}, \citenamefont {Luo}, \citenamefont {Majorana}, \citenamefont {Marchio}, \citenamefont {Michimura}, \citenamefont {Mio}, \citenamefont {Miyakawa}, \citenamefont {Miyamoto}, \citenamefont {Miyazaki}, \citenamefont {Miyo}, \citenamefont {Miyoki}, \citenamefont {Mori}, \citenamefont {Morisaki}, \citenamefont {Moriwaki}, \citenamefont {Nagano}, \citenamefont {Nagano}, \citenamefont {Nakamura}, \citenamefont {Nakano}, \citenamefont {Nakano}, \citenamefont {Nakashima},
  \citenamefont {Nakayama}, \citenamefont {Narikawa}, \citenamefont {Naticchioni}, \citenamefont {Negishi}, \citenamefont {Nguyen~Quynh}, \citenamefont {Ni}, \citenamefont {Nishizawa}, \citenamefont {Nozaki}, \citenamefont {Obuchi}, \citenamefont {Ogaki}, \citenamefont {Oh}, \citenamefont {Oh}, \citenamefont {Oh}, \citenamefont {Ohashi}, \citenamefont {Ohishi}, \citenamefont {Ohkawa}, \citenamefont {Ohta}, \citenamefont {Okutani}, \citenamefont {Okutomi}, \citenamefont {Oohara}, \citenamefont {Ooi}, \citenamefont {Oshino}, \citenamefont {Otabe}, \citenamefont {Pan}, \citenamefont {Pang}, \citenamefont {Parisi}, \citenamefont {Park}, \citenamefont {Peña~Arellano}, \citenamefont {Pinto}, \citenamefont {Sago}, \citenamefont {Saito}, \citenamefont {Saito}, \citenamefont {Sakai}, \citenamefont {Sakai}, \citenamefont {Sakuno}, \citenamefont {Sato}, \citenamefont {Sato}, \citenamefont {Sawada}, \citenamefont {Sekiguchi}, \citenamefont {Sekiguchi}, \citenamefont {Shao}, \citenamefont {Shibagaki}, \citenamefont
  {Shimizu}, \citenamefont {Shimoda}, \citenamefont {Shimode}, \citenamefont {Shinkai}, \citenamefont {Shishido}, \citenamefont {Shoda}, \citenamefont {Somiya}, \citenamefont {Son}, \citenamefont {Sotani}, \citenamefont {Sugimoto}, \citenamefont {Suresh}, \citenamefont {Suzuki}, \citenamefont {Suzuki}, \citenamefont {Tagoshi}, \citenamefont {Takahashi}, \citenamefont {Takahashi}, \citenamefont {Takamori}, \citenamefont {Takano}, \citenamefont {Takeda}, \citenamefont {Takeda}, \citenamefont {Tanaka}, \citenamefont {Tanaka}, \citenamefont {Tanaka}, \citenamefont {Tanaka}, \citenamefont {Tanaka}, \citenamefont {Tanioka}, \citenamefont {Tapia San~Martin}, \citenamefont {Telada}, \citenamefont {Tomaru}, \citenamefont {Tomigami}, \citenamefont {Tomura}, \citenamefont {Travasso}, \citenamefont {Trozzo}, \citenamefont {Tsang}, \citenamefont {Tsao}, \citenamefont {Tsubono}, \citenamefont {Tsuchida}, \citenamefont {Tsutsui}, \citenamefont {Tsuzuki}, \citenamefont {Tuyenbayev}, \citenamefont {Uchikata}, \citenamefont
  {Uchiyama}, \citenamefont {Ueda}, \citenamefont {Uehara}, \citenamefont {Ueno}, \citenamefont {Ueshima}, \citenamefont {Uraguchi}, \citenamefont {Ushiba}, \citenamefont {van Putten}, \citenamefont {Vocca}, \citenamefont {Wang}, \citenamefont {Washimi}, \citenamefont {Wu}, \citenamefont {Wu}, \citenamefont {Wu}, \citenamefont {Xu}, \citenamefont {Yamada}, \citenamefont {Yamamoto}, \citenamefont {Yamamoto}, \citenamefont {Yamamoto}, \citenamefont {Yamashita}, \citenamefont {Yamazaki}, \citenamefont {Yang}, \citenamefont {Yokogawa}, \citenamefont {Yokoyama}, \citenamefont {Yokozawa}, \citenamefont {Yoshioka}, \citenamefont {Yuzurihara}, \citenamefont {Zeidler}, \citenamefont {Zhan}, \citenamefont {Zhang}, \citenamefont {Zhao},\ and\ \citenamefont {Zhu}}]{akutsu_vibration_2021}%
  \BibitemOpen
  \bibfield  {author} {Akutsu \emph {et~al.},\ }\href {\doibase 10.1088/1361-6382/abd922} {\bibfield  {journal} {\bibinfo  {journal} {Classical and Quantum Gravity}\ }\textbf {\bibinfo {volume} {38}},\ \bibinfo {pages} {065011} (\bibinfo {year} {2021})},\ \bibinfo {note} {publisher: IOP Publishing}\BibitemShut {NoStop}%
\bibitem [{\citenamefont {Koroveshi}\ \emph {et~al.}(2023)\citenamefont {Koroveshi}, \citenamefont {Busch}, \citenamefont {Majorana}, \citenamefont {Puppo}, \citenamefont {Rapagnani}, \citenamefont {Ricci}, \citenamefont {Ruggi},\ and\ \citenamefont {Grohmann}}]{koroveshi_cryogenic_2023}%
  \BibitemOpen
  \bibfield  {author} {Koroveshi \emph {et~al.},\ }\href {\doibase 10.1103/physrevd.108.123009} {\bibfield  {journal} {\bibinfo  {journal} {Physical Review D}\ }\textbf {\bibinfo {volume} {108}} (\bibinfo {year} {2023}),\ 10.1103/physrevd.108.123009},\ \bibinfo {note} {publisher: American Physical Society (APS)}\BibitemShut {NoStop}%
\bibitem [{\citenamefont {Utina}\ \emph {et~al.}(2022)\citenamefont {Utina}, \citenamefont {Amato}, \citenamefont {Arends}, \citenamefont {Arina}, \citenamefont {de~Baar}, \citenamefont {Baars}, \citenamefont {Baer}, \citenamefont {van Bakel}, \citenamefont {Beaumont}, \citenamefont {Bertolini}, \citenamefont {van Beuzekom}, \citenamefont {Biersteker}, \citenamefont {Binetti}, \citenamefont {ter Brake}, \citenamefont {Bruno}, \citenamefont {Bryant}, \citenamefont {Bulten}, \citenamefont {Busch}, \citenamefont {Cebeci}, \citenamefont {Collette}, \citenamefont {Cooper}, \citenamefont {Cornelissen}, \citenamefont {Cuijpers}, \citenamefont {van Dael}, \citenamefont {Danilishin}, \citenamefont {Diksha}, \citenamefont {van Doesburg}, \citenamefont {Doets}, \citenamefont {Elsinga}, \citenamefont {Erends}, \citenamefont {van Erps}, \citenamefont {Freise}, \citenamefont {Frenaij}, \citenamefont {Garcia}, \citenamefont {Giesberts}, \citenamefont {Grohmann}, \citenamefont {Van~Haevermaet}, \citenamefont {Heijnen},
  \citenamefont {van Heijningen}, \citenamefont {Hennes}, \citenamefont {Hennig}, \citenamefont {Hennig}, \citenamefont {Hertog}, \citenamefont {Hild}, \citenamefont {Hoffmann}, \citenamefont {Hoft}, \citenamefont {Hopman}, \citenamefont {Hoyland}, \citenamefont {Iandolo}, \citenamefont {Ietswaard}, \citenamefont {Jamshidi}, \citenamefont {Jansweijer}, \citenamefont {Jones}, \citenamefont {Jones}, \citenamefont {Knust}, \citenamefont {Koekoek}, \citenamefont {Koroveshi}, \citenamefont {Kortekaas}, \citenamefont {Koushik}, \citenamefont {Kraan}, \citenamefont {van~de Kraats}, \citenamefont {Kranzhoff}, \citenamefont {Kuijer}, \citenamefont {Kukkadapu}, \citenamefont {Lam}, \citenamefont {Letendre}, \citenamefont {Li}, \citenamefont {Limburg}, \citenamefont {Linde}, \citenamefont {Locquet}, \citenamefont {Loosen}, \citenamefont {Lueck}, \citenamefont {Martínez}, \citenamefont {Masserot}, \citenamefont {Meylahn}, \citenamefont {Molenaar}, \citenamefont {Mow-Lowry}, \citenamefont {Mundet}, \citenamefont
  {Munneke}, \citenamefont {van Nieuwland}, \citenamefont {Pacaud}, \citenamefont {Pascucci}, \citenamefont {Petit}, \citenamefont {Van~Ranst}, \citenamefont {Raskin}, \citenamefont {Recaman}, \citenamefont {van Remortel}, \citenamefont {Rolland}, \citenamefont {de~Roo}, \citenamefont {Roose}, \citenamefont {Rosier}, \citenamefont {Ryckbosch}, \citenamefont {Schouteden}, \citenamefont {Sevrin}, \citenamefont {Sider}, \citenamefont {Singha}, \citenamefont {Spagnuolo}, \citenamefont {Stahl}, \citenamefont {Steinlechner}, \citenamefont {Steinlechner}, \citenamefont {Swinkels}, \citenamefont {Szilasi}, \citenamefont {Tacca}, \citenamefont {Thienpont}, \citenamefont {Vecchio}, \citenamefont {Verkooijen}, \citenamefont {Vermeer}, \citenamefont {Vervaeke}, \citenamefont {Visser}, \citenamefont {Walet}, \citenamefont {Werneke}, \citenamefont {Westhofen}, \citenamefont {Willke}, \citenamefont {Xhahi},\ and\ \citenamefont {Zhang}}]{utina_etpathfinder_2022}%
  \BibitemOpen
  \bibfield  {author} {Utina \emph {et~al.},\ }\href {\doibase 10.1088/1361-6382/ac8fdb} {\bibfield  {journal} {\bibinfo  {journal} {Classical and Quantum Gravity}\ }\textbf {\bibinfo {volume} {39}},\ \bibinfo {pages} {215008} (\bibinfo {year} {2022})},\ \bibinfo {note} {publisher: IOP Publishing}\BibitemShut {NoStop}%
\bibitem [{noa(2025)}]{noauthor_mineral_2025}%
  \BibitemOpen
  \href {https://www.usgs.gov/centers/national-minerals-information-center/mineral-commodity-summaries} {\enquote {\bibinfo {title} {Mineral {Commodity} {Summaries} {\textbar} {U}.{S}. {Geological} {Survey}},}\ } (\bibinfo {year} {2025})\BibitemShut {NoStop}%
\bibitem [{\citenamefont {Bertolini}\ \emph {et~al.}(1999)\citenamefont {Bertolini}, \citenamefont {Cella}, \citenamefont {DeSalvo},\ and\ \citenamefont {Sannibale}}]{bertolini_seismic_1999}%
  \BibitemOpen
  \bibfield  {author} {Bertolini \emph {et~al.},\ }\href {\doibase 10.1016/S0168-9002(99)00554-9} {\bibfield  {journal} {\bibinfo  {journal} {Nuclear Instruments and Methods in Physics Research Section A: Accelerators, Spectrometers, Detectors and Associated Equipment}\ }\textbf {\bibinfo {volume} {435}},\ \bibinfo {pages} {475} (\bibinfo {year} {1999})}\BibitemShut {NoStop}%
\bibitem [{\citenamefont {Beker}\ \emph {et~al.}(2012)\citenamefont {Beker}, \citenamefont {Blom}, \citenamefont {van~den Brand}, \citenamefont {Bulten}, \citenamefont {Hennes},\ and\ \citenamefont {Rabeling}}]{beker_seismic_2012}%
  \BibitemOpen
  \bibfield  {author} {Beker \emph {et~al.},\ }\href {\doibase 10.1016/j.phpro.2012.03.741} {\bibfield  {journal} {\bibinfo  {journal} {Physics Procedia}\ }\bibinfo {series} {Proceedings of the 2nd {International} {Conference} on {Technology} and {Instrumentation} in {Particle} {Physics} ({TIPP} 2011)},\ \textbf {\bibinfo {volume} {37}},\ \bibinfo {pages} {1389} (\bibinfo {year} {2012})}\BibitemShut {NoStop}%
\bibitem [{\citenamefont {Kumar}\ \emph {et~al.}(2016)\citenamefont {Kumar}, \citenamefont {Chen}, \citenamefont {Hagiwara}, \citenamefont {Kajita}, \citenamefont {Miyamoto}, \citenamefont {Suzuki}, \citenamefont {Sakakibara}, \citenamefont {Tanaka}, \citenamefont {Yamamoto},\ and\ \citenamefont {Tomaru}}]{kumar_status_2016}%
  \BibitemOpen
  \bibfield  {author} {Kumar \emph {et~al.},\ }\href {\doibase 10.1088/1742-6596/716/1/012017} {\bibfield  {journal} {\bibinfo  {journal} {Journal of Physics: Conference Series}\ }\textbf {\bibinfo {volume} {716}},\ \bibinfo {pages} {012017} (\bibinfo {year} {2016})},\ \bibinfo {note} {publisher: IOP Publishing}\BibitemShut {NoStop}%
\bibitem [{\citenamefont {Eßer}\ \emph {et~al.}(2024)\citenamefont {Eßer}, \citenamefont {Pratzer}, \citenamefont {Frömming}, \citenamefont {Duffhauß}, \citenamefont {Bhaskar}, \citenamefont {Krzyzowski},\ and\ \citenamefont {Morgenstern}}]{eser_ultra-high_2024}%
  \BibitemOpen
  \bibfield  {author} {Eßer \emph {et~al.},\ }\href {\doibase 10.1063/5.0230892} {\bibfield  {journal} {\bibinfo  {journal} {Review of Scientific Instruments}\ }\textbf {\bibinfo {volume} {95}},\ \bibinfo {pages} {123703} (\bibinfo {year} {2024})}\BibitemShut {NoStop}%
\bibitem [{\citenamefont {Deshpande}\ \emph {et~al.}(2025)\citenamefont {Deshpande}, \citenamefont {Ionescu}, \citenamefont {Miller}, \citenamefont {Wang}, \citenamefont {Gabrielse}, \citenamefont {Geraci},\ and\ \citenamefont {Kovachy}}]{deshpande_demonstration_2025}%
  \BibitemOpen
  \bibfield  {author} {Deshpande \emph {et~al.},\ }\href {\doibase 10.48550/arXiv.2412.20623} {\enquote {\bibinfo {title} {Demonstration that {Differential} {Length} {Changes} of {Optical} {Cavities} are a {Sensitive} {Probe} for {Ultralight} {Dark} {Matter}},}\ } (\bibinfo {year} {2025}),\ \bibinfo {note} {arXiv:2412.20623 [hep-ex]}\BibitemShut {NoStop}%
\bibitem [{\citenamefont {Stochino}\ \emph {et~al.}(2007)\citenamefont {Stochino}, \citenamefont {DeSalvo}, \citenamefont {Huang},\ and\ \citenamefont {Sannibale}}]{stochino_improvement_2007}%
  \BibitemOpen
  \bibfield  {author} {Stochino \emph {et~al.},\ }\href {\doibase 10.1016/j.nima.2007.06.029} {\bibfield  {journal} {\bibinfo  {journal} {Nuclear Instruments and Methods in Physics Research Section A: Accelerators, Spectrometers, Detectors and Associated Equipment}\ }\textbf {\bibinfo {volume} {580}},\ \bibinfo {pages} {1559} (\bibinfo {year} {2007})}\BibitemShut {NoStop}%
\bibitem [{\citenamefont {Blom}(2015)}]{blom_seismic_2015}%
  \BibitemOpen
  \bibfield  {author} {Blom,\ }\href {https://research.vu.nl/en/publications/seismic-attenuation-for-advanced-virgo-vibration-isolation-for-th} {\bibfield  {journal} {\bibinfo  {journal} {PhD Thesis}\ } (\bibinfo {year} {2015})}\BibitemShut {NoStop}%
\bibitem [{\citenamefont {Cella}\ \emph {et~al.}(2005)\citenamefont {Cella}, \citenamefont {Sannibale}, \citenamefont {DeSalvo}, \citenamefont {Márka},\ and\ \citenamefont {Takamori}}]{cella_monolithic_2005}%
  \BibitemOpen
  \bibfield  {author} {Cella \emph {et~al.},\ }\href {\doibase 10.1016/j.nima.2004.10.042} {\bibfield  {journal} {\bibinfo  {journal} {Nuclear Instruments and Methods in Physics Research Section A: Accelerators, Spectrometers, Detectors and Associated Equipment}\ }\textbf {\bibinfo {volume} {540}},\ \bibinfo {pages} {502} (\bibinfo {year} {2005})}\BibitemShut {NoStop}%
\bibitem [{Note1()}]{Note1}%
  \BibitemOpen
  \bibinfo {note} {The drawings will be uploaded online soon.}\BibitemShut {Stop}%
\bibitem [{\citenamefont {Naimon}, \citenamefont {Weston},\ and\ \citenamefont {Ledbetter}(1974)}]{naimon_elastic_1974}%
  \BibitemOpen
  \bibfield  {author} {Naimon \emph {et~al.},\ }\href {\doibase 10.1016/0011-2275(74)90223-9} {\bibfield  {journal} {\bibinfo  {journal} {Cryogenics}\ }\textbf {\bibinfo {volume} {14}},\ \bibinfo {pages} {246} (\bibinfo {year} {1974})}\BibitemShut {NoStop}%
\bibitem [{\citenamefont {Marquardt}, \citenamefont {Le},\ and\ \citenamefont {Radebaugh}(2001)}]{marquardt_cryogenic_2001}%
  \BibitemOpen
  \bibfield  {author} {Marquardt \emph {et~al.},\ }\href {https://www.nist.gov/publications/cryogenic-material-properties-database} {\bibfield  {journal} {\bibinfo  {journal} {NIST}\ ,\ \bibinfo {pages} {681}} (\bibinfo {year} {2001})},\ \bibinfo {note} {last Modified: 2017-02-19T20:02-05:00 Publisher: E D. Marquardt, J P. Le, Ray Radebaugh}\BibitemShut {NoStop}%
\bibitem [{\citenamefont {Gilyazov}\ \emph {et~al.}(2023)\citenamefont {Gilyazov}, \citenamefont {Sibgatullin}, \citenamefont {Achmerov}, \citenamefont {Plotnikova},\ and\ \citenamefont {Salakhov}}]{gilyazov_development_2023}%
  \BibitemOpen
  \bibfield  {author} {Gilyazov \emph {et~al.},\ }\href {\doibase 10.1007/s10553-023-01503-x} {\bibfield  {journal} {\bibinfo  {journal} {Chemistry and Technology of Fuels and Oils}\ }\textbf {\bibinfo {volume} {59}},\ \bibinfo {pages} {58} (\bibinfo {year} {2023})}\BibitemShut {NoStop}%
\bibitem [{\citenamefont {Allan}\ \emph {et~al.}(2025)\citenamefont {Allan}, \citenamefont {Bas}, \citenamefont {Cornelis}, \citenamefont {Milan},\ and\ \citenamefont {Tjerk}}]{allan_system_2025}%
  \BibitemOpen
Allan \emph {et~al.}\bibfield  {author} {,\ }NL2035605B1,\ System for reducing vibrations\href {https://patents.google.com/patent/NL2035605B1/nl?q=(%22milan+P.+Allan%22)&oq=%22milan+P.+Allan%22} {\  (\bibinfo {year} {2025})}\BibitemShut {NoStop}%
\end{thebibliography}%
\end{document}